\documentclass[prd,showpacs,letterpaper,twocolumn,nofootinbib,longbibliography]{revtex4-1}%
\usepackage{graphicx}
\usepackage{bm}
\usepackage{epsf}
\usepackage{rotating}
\usepackage{epsfig,graphics,rotate,color}
\usepackage{wrapfig}
\usepackage{amssymb}
\usepackage{amsmath}
\usepackage{amsfonts}
\usepackage{braket}
\usepackage[utf8]{inputenc}
\usepackage[dvipsnames]{xcolor}
\usepackage[colorlinks=true,citecolor=blue,linkcolor=blue]{hyperref}
\usepackage{cleveref}
\usepackage{mathbbol}
\usepackage{tabularx}
\usepackage{listings}

\usepackage{placeins} 

\usepackage{tikz}
\usepackage[compat=1.1.0]{tikz-feynman}
\usepackage{dsfont}
\usepackage{verbatim}

\usepackage{empheq}
 
\newlength\dlf  

\usepackage{array,hhline,dcolumn}%
\providecommand{\U}[1]{\protect\rule{.1in}{.1in}}

\newcommand{\ttf}{\ttfamily}
\definecolor{kyelloworange}   {RGB}{255, 210,  110}

\begin{document}

\title{Exploring a Non-Minimal Sterile Neutrino Model Involving Decay
  at IceCube and Beyond}
\author{Z. Moss}
\email{zander@caltech.edu}
\affiliation{Dept.~of Physics, Massachusetts Institute of Technology, Cambridge, MA 02139, USA}
\author{M.H. Moulai}
\email{marjon@mit.edu}
\affiliation{Dept.~of Physics, Massachusetts Institute of Technology, Cambridge, MA 02139, USA}
\author{C.A. Arg\"uelles}
\email{caad@mit.edu}
\affiliation{Dept.~of Physics, Massachusetts Institute of Technology, Cambridge, MA 02139, USA}
\author{J.M. Conrad}
\email{conrad@mit.edu}
\affiliation{Dept.~of Physics, Massachusetts Institute of Technology, Cambridge, MA 02139, USA}

\begin{abstract}
We study the phenomenology of neutrino decay together with neutrino oscillations in the context of eV-scale sterile neutrinos. We review the formalism of visible neutrino decay in which one of the decay products is a neutrino that potentially can be observed. We apply the formalism developed for decay to the recent sterile neutrino search performed by IceCube with TeV neutrinos. We show that for $\nu_4$ lifetime $\tau_4/m_4 \lesssim 10^{-16} {\rm eV^{-1}s}$, the interpretation of the high-energy IceCube analysis can be significantly changed.
\end{abstract}


\maketitle

\section{Introduction\label{sec:intro}}
Oscillations between the three active flavors of neutrinos, $\nu_e$, $\nu_\mu$, and $\nu_\tau$, have been definitively observed~\cite{PhysRevLett.81.1562, PhysRevLett.89.011301}. The mixing angles and mass-squared differences that describe these oscillations have been well measured~\cite{nufit}, but the CP violating phase value remains unknown. Nevertheless, there remain anomalies in accelerator~\cite{Athanassopoulos:1996jb, Aguilar:2001ty, Aguilar-Arevalo:2012fmn}, reactor~\cite{Mention:2011rk}, and radioactive source~\cite{Bahcall:1994bq} experiments that do not fit this model well. These anomalies are often explained by introducing a new neutrino state, $\nu_s$, that does not participate in the Standard Model (SM) weak interactions, hence the name sterile neutrino.  As among the active neutrinos, mass mixing induces new flavor transitions between the sterile and the active neutrino flavors. The mass spectrum of the minimal sterile neutrino model contains three mostly active, light neutrinos, $\nu_1$, $\nu_2$, and $\nu_3$, and one mostly sterile neutrino, $\nu_4$. In the minimal 3+1 sterile neutrino model, $\nu_4$ is much heavier than the other mass eigenstates.

Searches for sterile neutrinos performed by the IceCube~\cite{TheIceCube:2016oqi, Aartsen:2017bap}, MINOS~\cite{Adamson:2011ku}, Super-Kamiokande~\cite{Abe:2014gda}, KARMEN~\cite{Armbruster:2002mp}, MiniBooNE/SciBooNE~\cite{Cheng:2012yy}, and CDHS~\cite{DYDAK1984281} collaborations have found null results. This has produced increasing tension between the favored regions of parameter space found in global fits to short baseline data and null results~\cite{Kopp:2013vaa, Collin:2016rao, Giunti:2011gz,Gariazzo:2017fdh}. This tension has led to the consideration of more complicated new physics scenarios. These include: keV fourth neutrino with decay~\cite{Ma:1999im, Ma:2001ip, Dib:2011jh, Gninenko:2010pr, Gninenko:2009ks}, 
three and four neutrinos with CPT violation~\cite{Strumia:2002fw, Murayama:2000hm, Barenboim:2002ah, GonzalezGarcia:2003jq, Barger:2003xm, Diaz:2010ft, Barenboim:2004wu}, five-neutrino oscillation~\cite{Sorel:2003hf}, quantum decoherence~\cite{Barenboim:2004wu, Farzan:2008zv}, Lorentz violation~\cite{deGouvea:2006qd, Diaz:2010ft}, sterile neutrinos in extra dimensions~\cite{Pas:2005rb, Carena:2017qhd, Hollenberg:2009ws}, neutrinos with varying mass~\cite{Zurek:2004vd, Kaplan:2004dq, Schwetz:2007cd}, muon decay with lepton number violation~\cite{Babu:2002ica, Armbruster:2003pq, Gaponenko:2004mi}, three twin-neutrinos \cite{Bai:2015ztj}, neutrino-antineutrino oscillations~\cite{Hollenberg:2009tr}, neutrino decay in the unparticle scenario~\cite{Li:2007kj}, CP violation from neutral heavy leptons~\cite{Nelson:2010hz}, and nonstandard interactions~\cite{Liao:2016reh, Akhmedov:2010vy, Papoulias:2016edm}. Here, we consider in detail the interplay of neutrino oscillation and neutrino decay in a 3+1 sterile neutrino model. 

Neutrino decay is predicted by the SM, but the rate is too small to be detected by present experimental searches. Neutrino decay via some new physics process may be important and has previously been considered in the context of: solar neutrinos~\cite{Berezhiani:1991vk,Joshipura:2002fb, Beacom:2002cb, Bandyopadhyay:2002qg, Choubey:2000an, Bandyopadhyay:2001ct}, atmospheric neutrinos~\cite{Barger:1998xk, Fogli:1999qt, Barger:1999bg, GonzalezGarcia:2008ru,Choubey:2017dyu, Choubey:2017eyg}, accelerator neutrinos~\cite{PalomaresRuiz:2005vf, PalomaresRuiz:2006ue, GonzalezGarcia:2008ru,Coloma:2017zpg}, supernova neutrinos~\cite{Lindner:2001th, Kachelriess:2000qc, Ando:2003ie, Fogli:2004gy}, cosmology~\cite{Serpico:2007pt, Hannestad:2005ex}, and high-energy astrophysical neutrinos~\cite{Maltoni:2008jr,Shoemaker:2015qul,Bustamante:2016ciw}. Although neutrino decay is an interesting theoretical proposition, it has not been observed and is very constrained for active neutrinos~\cite{Joshipura:2002fb, Hannestad:2005ex, Gomes:2014yua, Abrahao:2015rba}. However, as we will discuss in detail in this work, constraints on the mostly sterile neutrino lifetime are weaker, and the IceCube results can be significantly altered, as we show in our main result.

The outline of this paper is as follows. In section~\ref{sec:decay_models} we introduce the neutrino decay model. In section~\ref{sec:osc_w_decay} we develop our framework for models of neutrino oscillation with neutrino decay for two scenarios. First we consider the case where the neutrino daughters are all Beyond the Standard Model (BSM) particles, which we term ''invisible decay". Secondly, we consider the case where one neutrino daughter is a lighter SM neutrino, which we refer to as ``visible decay". In section~\ref{sec:application_icecube} we illustrate our model in the IceCube experiment using atmospheric neutrino data. Finally, in section~\ref{sec:conclusions}, we provide concluding remarks.

\section{Neutrino decay models\label{sec:decay_models}}

\begin{figure}[tbp!]
    \centering
    \begin{tikzpicture}
        \fill[kyelloworange]   (2cm,0)  circle (0.36cm);
        \begin{feynman}
            \vertex (qi) {$\nu_i$};
            \vertex [right=2cm of qi] (c) {$ig_{ij}$} ;
            \vertex [below right=2cm of c] (qf1) {$\nu_j$};
            \vertex [above right=2cm of c] (pf) {$\phi$};
            \diagram*{
                (qi) -- [fermion] 
                (c) --  [scalar] (pf),
                (c) -- [fermion] (qf1),
                };
        \end{feynman}
    \end{tikzpicture}
    \caption{Decay of a massive neutrino state to a lighter neutrino mass state and a scalar $\phi$.}
    \label{fig:nudecay}
\end{figure}
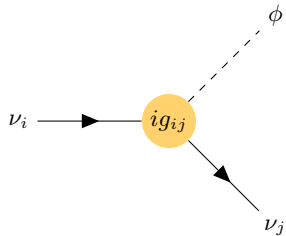

In this paper, we are interested in neutrino decay via new interactions. The simplest cases are those in which the neutrino decays into two particles. These are expected to be dominant over decays into three or more particles, as these involve high dimensional operators~\cite{Schechter:1981cv}. Fig.~\ref{fig:nudecay} shows a diagram of this process. We can organize the two-daughter processes into (I) invisible and (II) visible decays. Note that in (I), if one of the new particles decays into neutrinos, then this can be reduced to an effective three-body decay, which we have chosen to neglect. From these statements, it follows that in (I) $\nu_i \to \phi \psi$ and in (II) $\nu_i \to \nu_j \phi$, where $\psi$ is a fermion and $\phi$ a boson. Due to it being phenomenologically more interesting, we will restrict our examples to the  visible decay scenario case, i.e., $\psi$ is one of the SM neutrinos, and consider $\phi$ to be either a scalar or pseudoscalar. 

\subsection{Lagrangians of simplified models\label{sec:simplify_models}}

In this work, we will use simplified models to introduce the aforementioned decay scenarios, as done, for example, in~\cite{Lindner:2001fx,Gago:2017zzy}. For simplicity, we write this section assuming that the neutrino is a Majorana field. This assumption can be relaxed by introducing right-handed neutrinos~\cite{Schechter:1981cv}, which would be related to Dirac mass terms and would give rise to a different phenomenology due to the lack of $\nu \to \bar \nu$ transitions and visible decay. With this assumption, the scalar-neutrino interaction is~\cite{Peskin:257493}
\begin{equation}
\mathcal{L}_{\rm int} =  \frac{g^s_{ij}}{2} \bar \nu^c_i \nu_j \phi + i \frac{g^p_{ij}}{2} \bar \nu^c_i \gamma_5 \nu_j \phi, 
\label{eq:int_lagrangian}
\end{equation}
where $g^s_{ij} (g^p_{ij})$ are the (pseudo)scalar couplings that control the transition strength from parent ($i$) to daughter ($j$): $\nu_i \to \nu_j \phi$. In other words, $g$ carries two indices: one for the parent mass eigenstate and another for the daughter neutrino mass eigenstate. Note that we assume $\hbar=c=1$ throughout, except where we explicitly restore constants to estimate decay lengths.

\subsection{Decay rates\label{sec:decay_rates}}

\begin{figure} [tbp!]
\centering
\includegraphics[width=\linewidth]{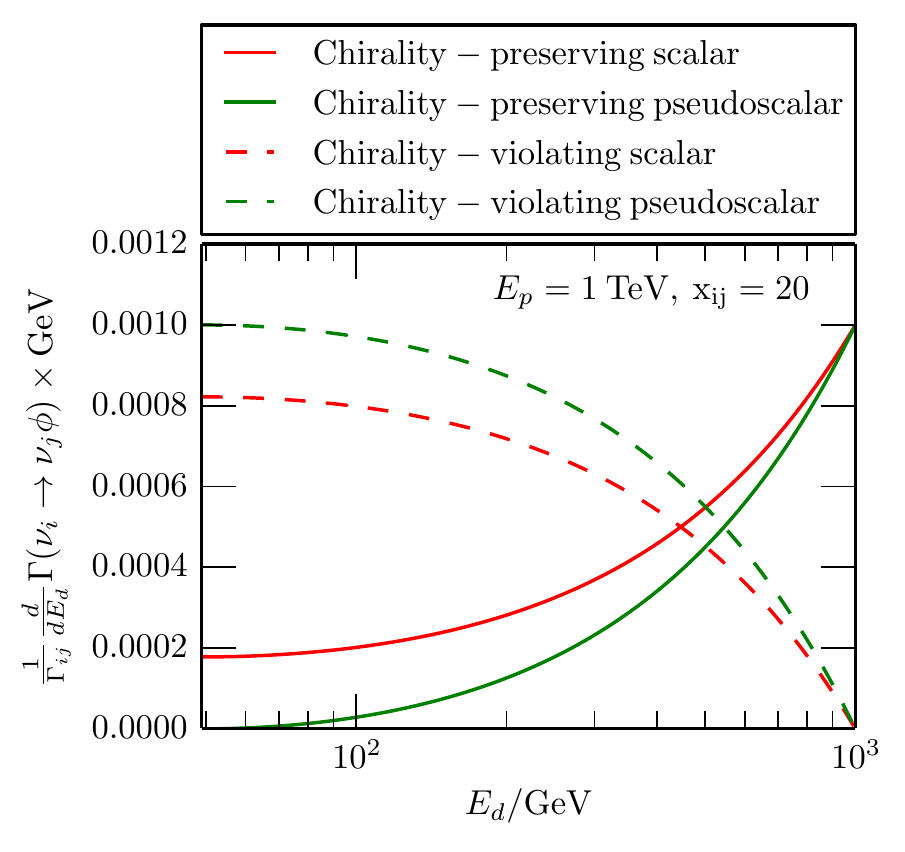}
\caption{Decay daughter energy distributions for chirality-preserving (solid) and chirality-violating (dashed) processes, for the scalar (red) and pseudoscalar (green) cases. In all these cases, we have set the parent energy to 1 TeV and $x_{ij}=20$. These energy distributions are given in equations Eq.~\eqref{eq:diff_rate_cpp} and Eq.~\eqref{eq:diff_rate_cvp}.}
\label{fig:decay_distribution}
\end{figure}

In the case where the boson is massless, our scenario is analogous to neutrino-Majoron decays. As in that case, there are two scenarios induced by Eq.~\eqref{eq:int_lagrangian}. The first one is a chirality-preserving process, whose partial decay rate is~\cite{doi:10.1142/S0217732390000354,Lindner:2001fx}
\begin{equation}
\Gamma (\nu_i^{(\pm)} \to \nu_j^{(\pm)} \phi) = \frac{m^2_i}{16\pi} \frac{1}{x_{ij} E_p} \left[ (g^s_{ij})^2 f(x_{ij}) + (g^p_{ij})^2 g(x_{ij}) \right],
\label{eq:gamma_chiral_preserving}
\end{equation}
where we have introduced the parent-to-daughter mass ratio, $x_{ij} = m_i/m_j >1$, and have labeled the energy of the parent neutrino, $\nu_i$, by $E_p$. The second one is a chirality-violating process
\begin{equation}
\Gamma (\nu_i^{(\pm)} \to \nu_j^{(\mp)} \phi) = \frac{m^2_i}{16\pi} \frac{1}{x_{ij} E_p} \left[ (g^s_{ij})^2 + (g^p_{ij})^2 \right] k (x_{ij}).
\label{eq:gamma_chiral_violating}
\end{equation}
In both cases, the auxiliary functions are given by~\cite{doi:10.1142/S0217732390000354}
\begin{subequations}
\begin{eqnarray}
f(x) &=& \frac{x}{2} + 2 + \frac{2}{x} \log x - \frac{2}{x^2} - \frac{1}{2 x^3},\\
g(x) &=& \frac{x}{2} - 2 + \frac{2}{x} \log x + \frac{2}{x^2} - \frac{1}{2 x^3},\\
k(x) &=& \frac{x}{2} - \frac{2}{x} \log x - \frac{1}{2 x^3},
\end{eqnarray}
\end{subequations}
where we have dropped the indices for clarity. If the lightest neutrino is massless, then for approximately eV sterile neutrinos, the smallest $x_{ij} \sim 10^2 \gg 1$, while if the lightest neutrino saturates the current kinematic limits, then the smallest $x_{ij} \sim 10$. In the limit of $m_i \gg m_j$, the partial decay rates given in Eq.~\eqref{eq:gamma_chiral_preserving} and Eq.~\eqref{eq:gamma_chiral_violating} are just~\cite{doi:10.1142/S0217732390000354}
\begin{equation}
\Gamma_{ij} = \left[(g^s_{ij})^2 + (g^p_{ij})^2\right] \frac{m^2_i}{32\pi E_p}.
\label{eq:simple_rate}
\end{equation}
In any case, the neutrinos that we are considering are relativistic, thus the products  of the decay will travel along the direction of the beam~\cite{Lindner:2001fx} and the relevant quantity is the energy distribution of the daughter in the lab frame. In the relativistic limit, the expression for this quantity simplifies if we assume either pure scalar ($g^p\equiv 0$) or pure pseudoscalar ($g^s\equiv 0$) cases. We will make this assumption throughout the rest of the paper, and in particular, we will assume a pure scalar couplings in our analysis. For the chirality-preserving process, the energy distribution takes the form~\cite{Gago:2017zzy}
\begin{equation}
\frac{1}{\Gamma_{ij}} \frac{d}{dE_d} \Gamma (\nu_i^{(\pm)} \to \nu_j^{(\pm)} \phi) = \frac{x_{ij}^2}{x_{ij}^2-1} \frac{1}{E^2_p E_d} \frac{(E_p \pm x_{ij} E_d)^2}{(x_{ij}  \pm 1)^2},
\label{eq:diff_rate_cpp}
\end{equation}
where $+$ corresponds to the pure scalar case and $-$ to the pure pseudoscalar one, and $E_d$ labels the energy of the daughter neutrino, $\nu_j$. For the chirality-violating process, the distribution takes the form
\begin{equation}
\frac{1}{\Gamma_{ij}} \frac{d}{dE_d} \Gamma (\nu_i^{(\pm)} \to \nu_j^{(\mp)} \phi) = \frac{x_{ij}^2}{x_{ij}^2-1} \frac{E_p -E_d}{E^2_p E_d} \frac{x^2_{ij}E_d - E_p}{(x_{ij}  \pm 1)^2}.
\label{eq:diff_rate_cvp}
\end{equation}
In both cases, due to kinematic constraints, the daughter energy is bounded to: $E_p/x_{ij}^2  \le E_d \le E_p$~\cite{Gago:2017zzy}. In the limit with $x_{ij} \gg 1$, which is the case for approximately eV sterile neutrinos, there is no distinction between the scalar and pseudoscalar scenarios. As can be seen from the energy dependence of these relationships, the chirality-preserving processes produce harder daughters, while the chirality-violating ones produce softer daughters (see Fig.~\ref{fig:decay_distribution}). It is important to note that the chirality-preserving processes dominate the visible decay due to the fact that the atmospheric neutrino flux is a steeply falling power law, $\phi(E)\sim E^{-3.7}$~\cite{PhysRevD.92.023004,Gaisser:2016uoy}; the soft daughters in the chirality-violating case are hidden below a large flux which reduces the visible case to the less interesting invisible decay scenario. The chirality-violating process in the Majorana context induces transitions between neutrinos and antineutrinos, but not if the term arises from Dirac neutrinos. In this latter case, the right-handed neutrinos are sterile and do not interact. Note that $x_{ij}$ diverges as $m_j \rightarrow 0$, but all decay rates and differential decay rates remain finite in this limit, so decays to massless daughters are well-defined in this framework. Finally, to easily compare with existing constraints on neutrino lifetime, we introduce the partial lifetime $
\tau_{ij} = 1/\Gamma_{ij}$.

\subsection{Existing constraints\label{sec:constraints}}

Constraints exist on the lifetime of neutrino mass eigenstates $\nu_i$ for $i<4$, i.e., for the active neutrinos. The constraints depend on the neutrino mass ordering as well as the absolute neutrino mass. In the normal ordering (NO), $\nu_3$ and $\nu_2$ are unstable and can decay to $\nu_1$, which is stable. On the other hand, in the inverted ordering (IO), $\nu_3$ is stable and the others unstable. Cosmology constrains both the sum of neutrino masses, such that $\sum_i m_i \lesssim 0.12$ at 95\% C.L.~\cite{Palanque-Delabrouille:2015pga}, and the radiative neutrino decay lifetime, to $\gtrsim 10^{19}$s \cite{Mirizzi:2007jd,Kim:2011ye}. However, constraints from cosmology on the presence of a fourth neutrino are model-dependent; assumptions required include the thermal history of the Universe and the influence of dark energy on the expansion history~\cite{KevAnnRev}. Moreover, sterile neutrino thermalization can be suppressed by a number of new physics scenarios~\cite{Kev2,Dasgupta:2013zpn,Hannestad:2013ana,Bento:2001xi,Chu:2006ua,Foot:1995bm,Gelmini:2004ah,Ho:2012br,Hamann:2011ge}; if sterile neutrinos do not thermalize, the bounds do not apply. 

On the other hand, given results from neutrino oscillation experiments~\cite{Gonzalez-Garcia:2014bfa}, all neutrino masses cannot be zero. The largest decay rate will be achieved when the stable neutrino is assumed to be massless, i.e., $m_1=0$ or $m_3=0$, for NO or IO, respectively. With this optimistic assumption, in IO the unstable neutrino masses cannot be less than $\sim 0.05~ {\rm eV}$ and in NO $m_2$ cannot be smaller than $\sim 0.008~ {\rm eV}$ and $m_3$ satisfies the same bound as for IO. The lifetime of a neutrino of mass $m_i$ is given by $\tau^{-1}_i = \sum_j \tau^{-1}_{ij}$, and the total decay rate in the lab frame is a function of $\tau_{i}/m_i$. Constraints on this quantity will depend on the mass ordering: for NO, $\nu_2$ decay is constrained by solar experiments, with $\tau_2/m_2 \gtrsim 7 \cdot 10^{-4} ~ {\rm s~ eV}^{-1}$~\cite{Berryman:2014qha}, and $\nu_3$ is limited by atmospheric and long-baseline experiments, with $\tau_3/m_3 \gtrsim 9 \cdot 10^{-11}~ {\rm s~ eV}^{-1}$ \cite{GonzalezGarcia:2008ru}; for IO, $\tau_1/m_1 \gtrsim 4 \cdot 10^{-3}~ {\rm s~ eV}^{-1}$ and $\tau_2/m_2 \gtrsim 7 \cdot 10^{-4}~ {\rm s~ eV}^{-1}$ are constrained by solar experiments~\cite{Berryman:2014qha}. In contrast, direct constraints on $\nu_4$ have not been set.

Constraints on neutrino decay can be obtained indirectly from measurements of meson decays~\cite{Barger:1981vd,Lessa:2007up,Pasquini:2015fjv}. These set strict limits on the neutrino decay process but are flavor-dependent; thus they must be used with care. For example, from kaon decay, the following combination is constrained~\cite{Barger:1981vd}
\begin{equation}
\sum_\alpha |g_{e\alpha}|^2 < 3 \times 10^{-5},
\end{equation}
where $\alpha$ runs over all neutrino flavors. Constraints from supernova 1987A are of similar order~\cite{Lessa:2007up}. These flavor couplings are related to the ones in Eq.~\eqref{eq:int_lagrangian} by 
\begin{equation}
g_{\alpha\beta} = \sum_{ij} g_{ij} U_{\alpha i}U_{\beta j}^*.
\label{eq:g_flavor}
\end{equation}
For simplicity, let's first consider the case where only one $g_{4j}$ is nonzero. In this case $g_{\alpha\beta} = g_{4j} U_{\alpha 4} U_{\beta j}^*$, where for short-baseline motivated sterile neutrinos $U_{\alpha 4}\sim O(0.1)$~\cite{Collin:2016aqd} and from standard neutrino measurements $U_{\beta j}\sim O(0.1)$ for $j<4$~\cite{nufit}, which implies that $g_{4j} \le O(0.1)$. For this size of coupling, an eV-scale sterile neutrino lifetime satisfies $\tau_4 > O(10)/{\rm eV}$. With this lifetime constraint, as we will see in a later example, no interesting interplay exists between oscillations and decay. This is due to the fact that the scales of oscillation and decay, which are given by $E/\Delta m_{4j}^2$ and $E \tau_4/m_4$, respectively, are very different. This implies that to have an interesting interplay, more than one of the $g_{4j}$ needs to be nonzero so that a cancellation can occur in Eq.~\eqref{eq:g_flavor}, leading to a decreased bound in $\tau_4$.

\begin{figure} [tbp!]
\centering
  \includegraphics[width=\linewidth]{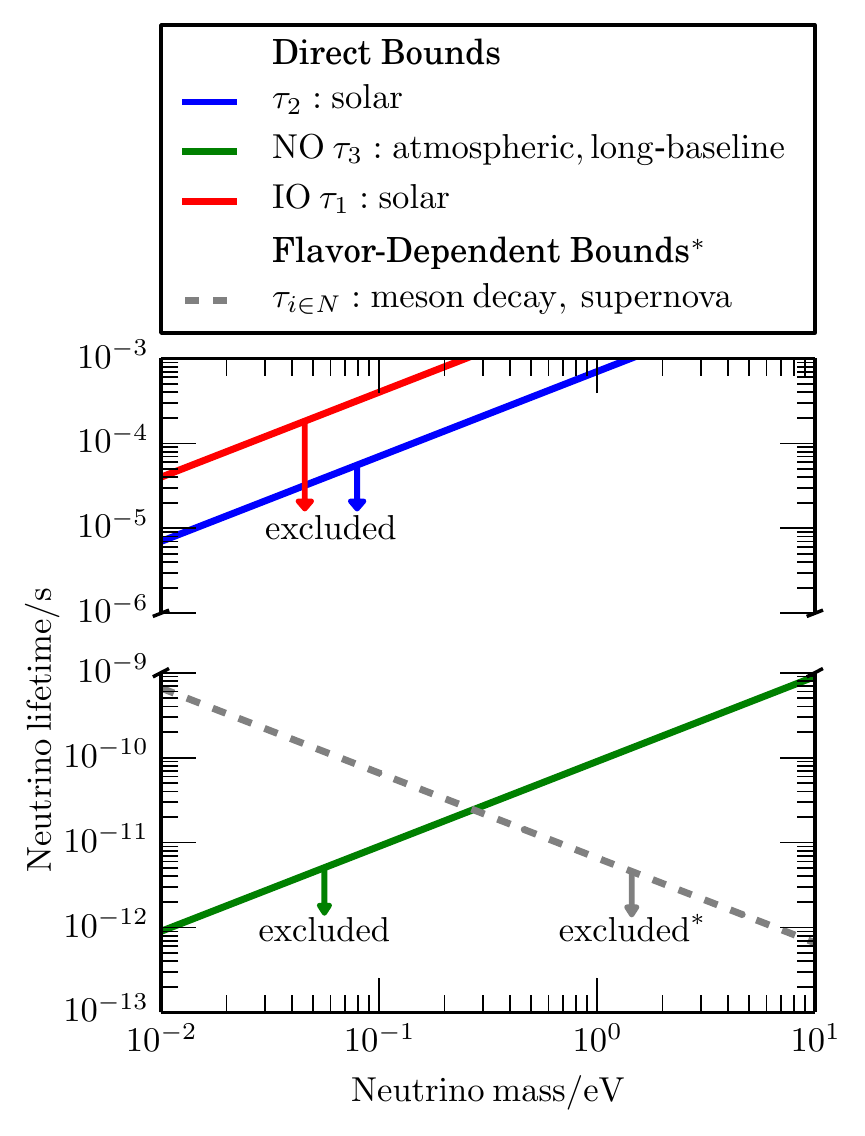}
  \caption{Direct bounds and flavor-dependent bounds on neutrino lifetimes as a function of the neutrino mass. Constraints are shown for both normal and inverted standard neutrino orderings; in both cases we assume a 3+1 sterile model. The flavor-dependent bounds are shown as dashed lines and may be relaxed by cancellations as discussed in the text.}
    \label{fig:constraints}
\end{figure}

We summarize the constraints discussed in this section in Fig.~\ref{fig:constraints} for both active mass states and the mostly sterile state, $\nu_4$.

\section{Neutrino oscillations and decay\label{sec:osc_w_decay}}

We will now develop a formalism that incorporates both oscillations and decay in a consistent way. We will follow the calculation in~\cite{GonzalezGarcia:2008ru,Berryman:2014yoa} for invisible decay, and for visible decay we follow~\cite{Lindner:2001fx,PalomaresRuiz:2006ue,Gago:2017zzy} but in the density matrix representation. We will work with a ($3+1$) model, which adds one sterile state to the three active neutrinos, although the model generalizes readily to ($3+N$) for generic $N$.

\subsection{Neutrino oscillations\label{sec:oscillations}}

We will begin with the most familiar model, which includes vacuum oscillations exclusively, ignoring decay and matter effects. We have
\begin{align}
H_{\mathrm{vacuum}} = \frac{\Delta M^2}{2E},
\end{align}
where $\Delta M^2$ is a diagonal matrix with entries $\left(\Delta M^2\right)_{ii} = \Delta m^2_{i1}$: the neutrino mass-squared splittings. Conjugation into the flavor basis with the PMNS matrix $U_\nu$ (appropriately extended to include the sterile state) mixes the mass states into flavor states as follows
\begin{align} \label{eq:basis_U}
\nu_\alpha = \sum U_{\nu;\alpha,i} \nu_i.
\end{align}
This mixing gives rise to vacuum oscillations when at least one of the mass splittings is non-zero.

We are interested in analyzing TeV-scale neutrinos in IceCube. At these energies, matter effects become important, so we introduce a term modeling neutrino scattering off of electrons and nucleons as an effective matter potential~\cite{Wolfenstein:1977ue,Mikheev:1986gs,Akhmedov:1999uz}. The flavor states are eigenstates of the corresponding Hamiltonian term $V_{\mathrm{matter}}$. In the flavor basis, this term takes the form $V_{\mathrm{matter}} = \mathrm{diag}(V_e,V_\mu, V_\tau, 0)$, where the fourth eigenvalue is zero because it corresponds to a sterile neutrino state. Thus the total Hamiltonian in the mass basis is given by
\begin{align}
H_0(E,l) = H_{\mathrm{vacuum}}(E)  + U_\nu V_{\mathrm{matter}}(E,l)U_\nu^\dagger ,
\end{align}
where the neutrino potentials depend on the electron ($N_e(l)$) and neutron ($N_n(l)$) number density profiles, and $l$ is the position of the neutrino ensemble along the baseline.

\subsection{Invisible neutrino decay\label{sec:invisible_decay}}

Modeling decay in an evolving neutrino system is not a trivial matter, because it involves the addition of daughter particles to the state space. In the visible decay case, we will treat these new states explicitly. In the case of invisible decay, the daughter states are irrelevant, and one can get around this added complexity by introducing an effective non-Hermitian Hamiltonian, $H$, which gives rise to a non-unitary evolution when restricted to the state space of the parent neutrinos. Intuitively, the non-unitarity of the associated time evolution operator corresponds to the loss of probability current from the parent state space into the daughter space. Following~\cite{Berryman:2014yoa} we can construct $H$ by the explicit addition of an anti-Hermitian term
\begin{align} \label{eq:ww_eff_ham}
H = H_0 -i\frac{1}{2}\Gamma,
\end{align}
where $\Gamma$ is a Hermitian operator that is diagonal in the mass basis, where it is just $\Gamma=  \mathrm{diag}(\Gamma_i(E), ...)$, where $\Gamma_i(E)$, the decay rate of the i$^{th}$ neutrino, is given by Eq.~\eqref{eq:simple_rate}. The factor of $\frac{1}{2}$ is necessary for the survival probability of a neutrino born in the $\nu_i$ mass eigenstate to follow the exponential decay formula \begin{equation}
P_{i \rightarrow i} = |\langle \nu_0 (l) | \nu_0 = \nu_i \rangle|^2 = e^{-\Gamma_i l}. 
\label{eq:exponential_decay_formula}
\end{equation}
In a two-flavor system, the active neutrino vacuum survival probability can written, under the assumption that the lighter of the two neutrinos is stable, as~\cite{GonzalezGarcia:2008ru} 
\begin{eqnarray}
P_{\alpha \rightarrow \alpha} = \cos^4 \theta &+& \frac{1}{2}e^{-\dfrac{\Gamma_2 l}{2}}\cos\left(\dfrac{\Delta m^2 l}{2E}\right) \sin^2 2\theta \nonumber \\
&+& e^{-\Gamma_2 l}\sin^4 \theta, 
\label{eq:decay_2_flavor}
\end{eqnarray}
where $\theta$ is the two-flavor mixing angle and $\Gamma_2$ is the decay rate of the heavier mass state with mass $m_2$. Equations \eqref{eq:exponential_decay_formula} and \eqref{eq:decay_2_flavor} are valid only in the case of invisible decay in vacuum and are included here for completeness. 

\subsection{Visible neutrino decay\label{sec:visible_decay}}

In this paper, we will concentrate on visible neutrino decay. In this case, Eq.~\eqref{eq:ww_eff_ham} cannot describe the full system evolution because it ignores the evolution of the daughter states. To treat the daughter states explicitly, we first need to extend our formalism from a single, monoenergetic neutrino state to a set of neutrino states indexed by the set of energies relevant to our analysis. We then promote each of these states to an ensemble of states, described by an $n\times n$ density matrix $\rho(E)$, where $n$ is the number of neutrino species under consideration. In the density matrix formalism, evolution due to the Hamiltonian given in Eq.~\eqref{eq:ww_eff_ham} can be written as
\begin{equation}
\frac{\partial \rho(E,l)}{\partial l} = -i [ H_0, \rho] - \frac{1}{2} \{ \Gamma, \rho \},
\label{eq:density_matrix_basic}
\end{equation}
where $l$ is the position of the neutrino ensemble along the baseline, square brackets indicate the commutator, and curly brackets indicate the anticommutator.  The factor of $\frac{1}{2}$ appears for a similar reason as in Eq.~\eqref{eq:ww_eff_ham}. This formalism has already been implemented in an efficient way in the context of neutrino oscillations in a package called {\ttf nuSQuIDS}~\cite{Delgado:2014kpa,nusquids}. The decay rate operator is given by
\begin{equation}
\Gamma(E) = \sum_i \Gamma_i(E) \Pi_i,
\label{eq:decay_op}
\end{equation}
where $\Pi_i$ is the projector to the i$^{th}$ mass eigenstate and $\Gamma_i$ is given by the sum of partial rates $\Gamma_{ij}$ over all daughter states $\nu_j$ lighter than the parent state $\nu_i$. The $\Gamma_{ij}$ are given by the sum of the chirality-preserving $\Gamma_{ij}$ from equation \eqref{eq:gamma_chiral_preserving} and the chirality-violating $\Gamma_{ij}$ from equation \eqref{eq:gamma_chiral_violating} in the case of Majorana neutrinos, or by partial rates $\Gamma_{ij}$ corresponding to purely chirality-violating processes in the case of Dirac neutrinos. 

Thus far, we have modeled only the loss of neutrinos using the $\Gamma$ term in Eq. \eqref{eq:density_matrix_basic}. The advantage of this extended formalism is that it can accommodate terms that generate transitions between neutrinos at different energies. We then complement the $\Gamma$ term describing the loss of neutrinos from an ensemble at one energy with a ``regeneration" term, $\mathcal{R}$, describing the appearance of these neutrinos in ensembles at lower energies. In this way, we keep track of the daughter states instead of neglecting them, and we account for their subsequent evolution.  

Let us assume that neutrinos are Majorana. For clarity, we will separate the contributions to regeneration into the chirality-preserving processes (CPP) and the chirality-violating processes (CVP). In the first case, we add the following term to the right-hand side Eq.~\eqref{eq:density_matrix_basic}
\begin{eqnarray}
\label{eq:r_cpp}
\mathcal{R}(E_d) &= \sum_{i,j} \int_{E_d}^{x_{ij}^2E_d} d E_p \bigg({\rm Tr}\left[\rho(E_p) \Pi_i(E_p)\right] \\ \nonumber
&\times\left(\frac{d\Gamma(E_p,E_d)}{dE_d}\right)_{ij}^{\rm CPP}  \Pi_j(E_d)\bigg),
\end{eqnarray}
where $E_p$ is the parent energy, $E_d$ is the daughter energy, ${\rm Tr}$ is the trace operation, and the differential decay rate is given in Eq.~\eqref{eq:diff_rate_cpp}. When this term is added to Eq.~\eqref{eq:density_matrix_basic}, we set $E_d=E$. We have an additional contribution from $\nu \to \bar\nu$ in the chirality-violating process, in which case we should add
\begin{eqnarray}
\label{eq:r_cvp}
\mathcal{R}(E_d) & \mathrel{{+}{=}} \sum_{i,j} \int_{E_d}^{x_{ij}^2E_d} d E_p \bigg({\rm Tr}\left[\bar\rho(E_p) \bar\Pi_i(E_p)\right]\\ \nonumber 
&\times \left(\frac{d\Gamma(E_p,E_d)}{dE_d}\right)_{ij}^{\rm CVP} \Pi_j(E_d)\bigg),
\end{eqnarray}
where $\bar\rho$ corresponds to antineutrinos and the differential rate is given by Eq.~\eqref{eq:diff_rate_cvp}. Because we consider  either purely scalar or purely pseudoscalar cases, the differential rates for CVP and CPP must be chosen accordingly when constructing $\mathcal{R}$. This boils down to getting the signs right in equations \eqref{eq:diff_rate_cpp} and \eqref{eq:diff_rate_cvp}. Similar equations hold for antineutrinos by replacing $\rho$ by $\bar \rho$ and changing the projectors $\Pi$ to $\bar\Pi$, and vice versa. Note that there is no regeneration in the Dirac case, because the decay is through a chirality-violating process that produces sterile daughters.
We have implemented this formalism in nuSQuIDS~\cite{Delgado:2014kpa,nusquids}; see Appendix A for details.

\section{Example of Application to IceCube\label{sec:application_icecube}}

The IceCube experiment is an ideal testing ground for the model presented here. The collaboration has released a 400~GeV to 20~ TeV single-year data set associated with a sterile neutrino search \cite{TheIceCube:2016oqi}, and an additional six years of data are available to the collaboration. In this section, we use the released data set to explore some of the nuances of including decays as additional phenomena in the IceCube search.  We will only consider the decay channel $\nu_4 \rightarrow \nu_3 \phi$, with the following $\nu_4$ lifetimes: (A) $\tau_4$ = 10 eV$^{-1}$, (B) $\tau_4$ = 1 eV$^{-1}$, (C) $\tau_4$ = 0.1 eV$^{-1}$, and $\tau_4$ = $\infty$ eV$^{-1}$, corresponding to no decay. For $\tau$ = 1 eV$^{-1}$, $\hbar c \tau \approx 0.2$ $\mu$m.

\subsection{The IceCube experiment and released data set\label{sec:icecube}}

The IceCube detector, located in the Antarctic ice below the South Pole station, observes Cherenkov radiation from interactions of neutrinos that have traversed the Earth. In the 400~GeV to 20~TeV energy range, muon neutrinos are primarily due to decays of kaons produced by cosmic rays impinging on the atmosphere~\cite{Gaisser:2016uoy}. This analysis makes use of through-going muons that are produced in charged-current $\nu_\mu$ interactions in the ice or bedrock below the detector. The muons produced by neutrinos in this energy range are above critical energy.  Hence, the muon energy can be determined from the stochastic light emission profile.  The direction of the muon can be determined through reconstruction of the Cherenkov-light time and spatial distribution and is expressed as an angle, $\theta_Z$,  with respect to the zenith.

The Cherenkov light is observed via digitial optical modules (DOMs), with photomultiplier tubes as the light sensors~\cite{Achterberg:2006md,Aartsen:2016nxy}.   The detector consists of 5,160 DOMs on 86 vertical strings. The intrastring DOM separation is 17 m and the interstring separation is approximately 125 m.
The energy resolution of the detector is $\sigma_{log_{10} (E_\mu/{\rm GeV})} \sim 0.5$ and the angular resolution, $\sigma_{\cos(\theta_Z)}$, varies from 0.005 to 0.015.

The released data set contains 20,145 well-reconstructed events, described in~\cite{TheIceCube:2016oqi, Jones:2015bya, Arguelles:2015}.  This release is associated with a sterile neutrino search with null results at 90\% CL~\cite{TheIceCube:2016oqi}.   The power of the sterile neutrino analysis arose from the matter effects that were expected for neutrinos that crossed the core and mantle~\cite{Nunokawa:2003ep}.   This was predicted to lead to an observable deficit of $\nu_\mu$ events in IceCube for parameters that were consistent with short baseline anomalies~\cite{Kopp:2013vaa, Collin:2016rao}.   The deficit was expected to be localized to  specific regions of $E_\mu$ and $\cos(\theta_Z)$, depending on the sterile neutrino parameter space.   As a result of the striking signature, IceCube was able to perform a powerful search.  The null result significantly changed the parameter-landscape for 3+1 searches~\cite{Collin:2016aqd}.

The released data provides energy and angle information for each data event.  It also provides Monte Carlo information on an event-by-event basis.    Information on handling of multiple error sources is provided.     Using this data set, one can reconstruct the IceCube search results well, as was demonstrated in~\cite{Collin:2016aqd}.    The appendix of that paper provides step-by-step instructions for use of the data release, which we follow here.

\subsection{IceCube oscillograms\label{sec:oscillogram}}

\begin{figure*}[t]
  \includegraphics[width=\textwidth]{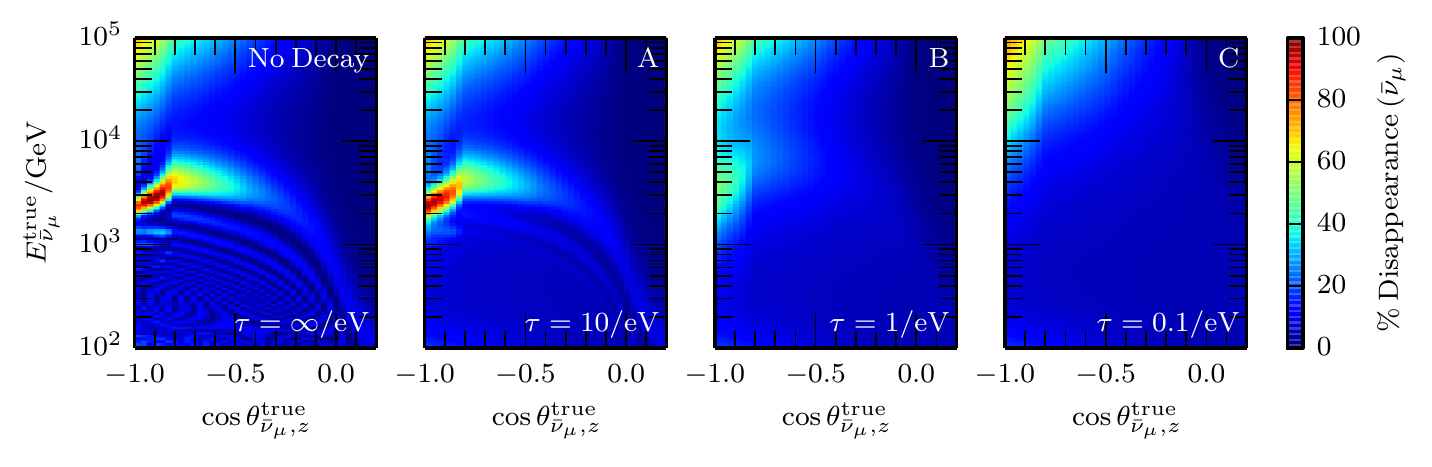}
  \caption{Disappearance probability for muon antineutrinos for sterile neutrino parameters $\Delta m_{41}^2 = 1~{\rm eV}^2$, 
$\sin ^2 2 \theta_{24} = 0.1$, and various lifetimes. Plots from left to right show the effect of neutrino decay with lifetime $\tau = \infty$, 10, 1, and 0.1 eV$^{-1}$. In all these plots, we assume the decay channel $\nu_4 \rightarrow \nu_3 \phi$. The visible Majorana scenario described in the text is assumed; oscillograms for invisible Majorana decay are within $\sim$ 10 \% of the ones shown here. For $\tau$ = 1 eV$^{-1}$, $\hbar c \tau \approx 0.2 \:\mu$m.}
\label{fig:oscillograms}
\end{figure*}

We refer to an ``oscillogram'' as a plot of the expected change in neutrino flux from creation in the atmosphere to arrival at IceCube, as a function of true neutrino zenith angle and true neutrino energy. Effects that may change the neutrino flux include oscillation, matter effects, absorption, and decay. The minimal 3+1 sterile neutrino model is parameterized by a mass-squared splitting, $\Delta m_{41}^2$, and a mixing angle, $\theta_{24}$. Fig.~\ref{fig:oscillograms} shows the shape effects in the $\bar{\nu}_\mu$ spectrum for the 3+1 sterile neutrino model with parameters $\Delta m_{41}^2 = 1 ~{\rm eV^2}$ and $\sin^2 2 \theta_{24} = 0.1$, for the four lifetimes listed previously. In Fig.~\ref{fig:oscillograms}, for no decay scenario as well as for examples points A and B, the depletion of $\bar{\nu}_\mu$s in the region $E_{\bar{\nu}_\mu}^{true} \sim$ 300~GeV and $\cos \theta_Z^{true} \sim$ -1.0 is due to matter effects. Decreasing the lifetime of $\nu_4$ decreases the magnitude of this feature and shifts its position. This is due to the fact that, as the $\nu_4$ lifetime becomes smaller, its decay length is smaller than the oscillation scale. In other words, the decay operation breaks the coherence of the system, projecting into the mostly active mass states, preventing the development of oscillation. However, the final flux is still different from the flux in the absence of oscillations. The depletion in the top-left corner of each plot in Fig.~\ref{fig:oscillograms} is due to absorption of high-energy neutrinos crossing the Earth.

\subsection{Data analysis and systematic uncertainties\label{sec:analysis}}

The data is binned in reconstructed energy proxy and zenith angle. We use a binned Poisson log-likelihood function, $\log\mathcal{L}$, for the data, and incorporate systematic uncertainties by means of nuisance parameters, $\vec\eta$, with Gaussian priors. For each lifetime considered and for each point in a fine, logarithmic grid of $[\sin^2 (2 \theta_{24}),\Delta m_{41}^2]$, $\log\mathcal{L}(\sin^2 (2 \theta_{24}),\Delta m_{41}^2,\vec\eta)$ is maximized with respect to a set of nuisance parameters.

Along with the ``conventional flux'' from pion and kaon decay, in principle a ``prompt'' contribution arises from the decays of heavier mesons, but it has yet to be observed~\cite{Enberg:2008te}. For this reason, the prompt flux is neglected~\cite{Fedynitch:2015zma}. The atmospheric neutrino flux is the sum of the neutrino component and the antineutrino component, where the neutrino component is parameterized as
\noindent
\begin{equation}
\phi_\nu^{\rm{atm}} = N_0 \mathcal{F}(\delta) \bigg(\phi^\pi_\nu + R_{K/\pi} \phi^K_\nu \bigg) \bigg( \frac{E_\nu}{E_0} \bigg)^{-\Delta \gamma},
\label{eq:atm_flux_equation}
\end{equation}
where $\phi_\nu^\pi$ and $\phi_\nu^K$ are the fluxes of neutrinos originating from decays of pions and kaons, respectively. The overall flux normalization, $N_0$; variations to the spectral index, $\Delta \gamma$; and the ratio of neutrinos originating from kaons and pions, $R_{K/\pi}$,  are nuisance parameters. The pivot point for the spectral index change, $E_0$, is at a midpoint energy such that changes to the spectral index do not dramatically change the overall flux normalization. The antineutrino flux is parametrized identically to the neutrino flux, up to a relative normalization factor, which is another nuisance parameter.  The cosmic ray and hadronic models used are Poly-gonato~\cite{Hoerandel:2002yg} and QGSJET-II-4~\cite{Ostapchenko:2010vb}, respectively. Uncertainty in the atmospheric density profile is accounted for in a linear parameterization, $\mathcal{F}(\delta)$~\cite{Jones:2015bya, Arguelles:2015}.

The DOM efficiency is a final nuisance parameter. Monte Carlo data sets corresponding to several discrete values of DOM efficiency are publicly available~\cite{IceDataRelease}. A piecewise linear interpolation was fit to them, allowing us to treat the DOM efficiency as a continuous nuisance parameter. Table~\ref{table:1} gives the prior values of the nuisance parameters and the best-fit values for the no-sterile-neutrino hypothesis.
\begin{table}[h!]
\centering
\begin{tabular}{l l l} 
 \hline
Parameter & Best Fit & Prior \\
\hline \hline
Flux normalization          & 1.3       & $1 \pm 0.4$\\
$\Delta \gamma$             & 0.006          & $0 \pm 0.05$ \\
K/$\pi$ ratio               & 1.1          & $1 \pm 0.1$\\
$\bar{\nu}/\nu$ ratio       & 1.0          & $1 \pm 0.05$ \\
DOM efficiency              & 1.0          & $1 \pm 1.0$\\
Atmospheric density shift    &  -0.01         & $0 \pm 0.0175$\\
\hline
\end{tabular}
\caption{Systematic uncertainties treated as nuisance parameters in this analysis. Parameters are described in the text. The best-fit values of the parameters for the null hypothesis (no sterile neutrino), as well as the Gaussian priors centers and widths, are given.}
\label{table:1}
\end{table}

\subsection{Results\label{sec:results}}

\begin{figure*}
  \includegraphics[width=\textwidth]{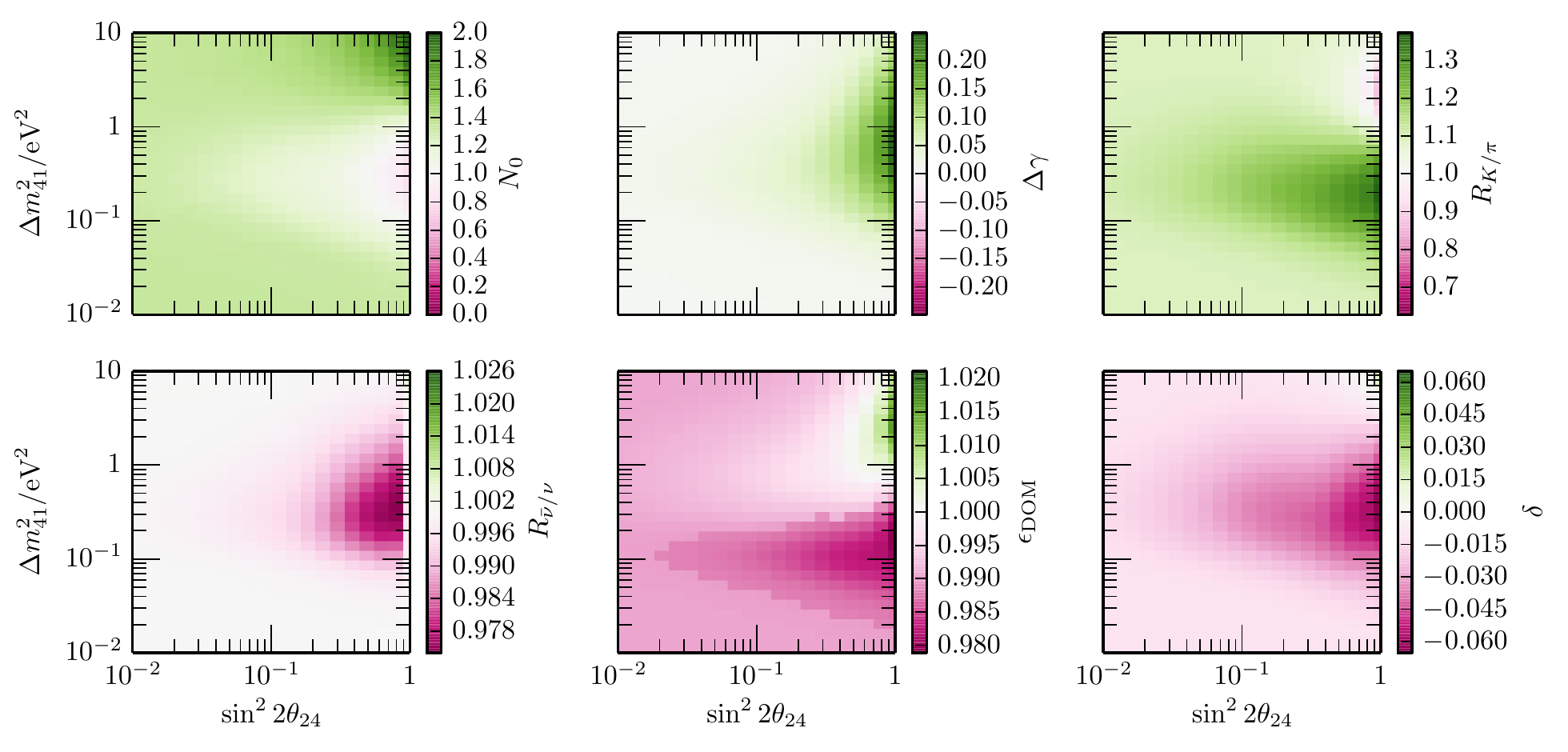}
  \caption{Minimized nuisance parameters over a scan of sterile parameters for a 3+1 sterile neutrino model with a stable $\nu_4$. The set of nuisance parameters are: (first row, from left to right) overall atmospheric flux normalization, change in cosmic ray spectral index, and ratio of atmospheric kaons to pions; (second row, from left to right) ratio of atmospheric muon antineutrinos to muon neutrinos, DOM efficiency, and atmospheric density uncertainty.}
  \label{fig:nuisance}
\end{figure*}

We have chosen a few specific parameters to illustrate the effect of this model. To demonstrate that our analysis technique is robust and properly implemented, we first perform the analysis without decay. Fig.~\ref{fig:nuisance} shows the best-fit nuisance parameters as functions of the sterile parameters.

\begin{figure*}
  \includegraphics[width=\textwidth]{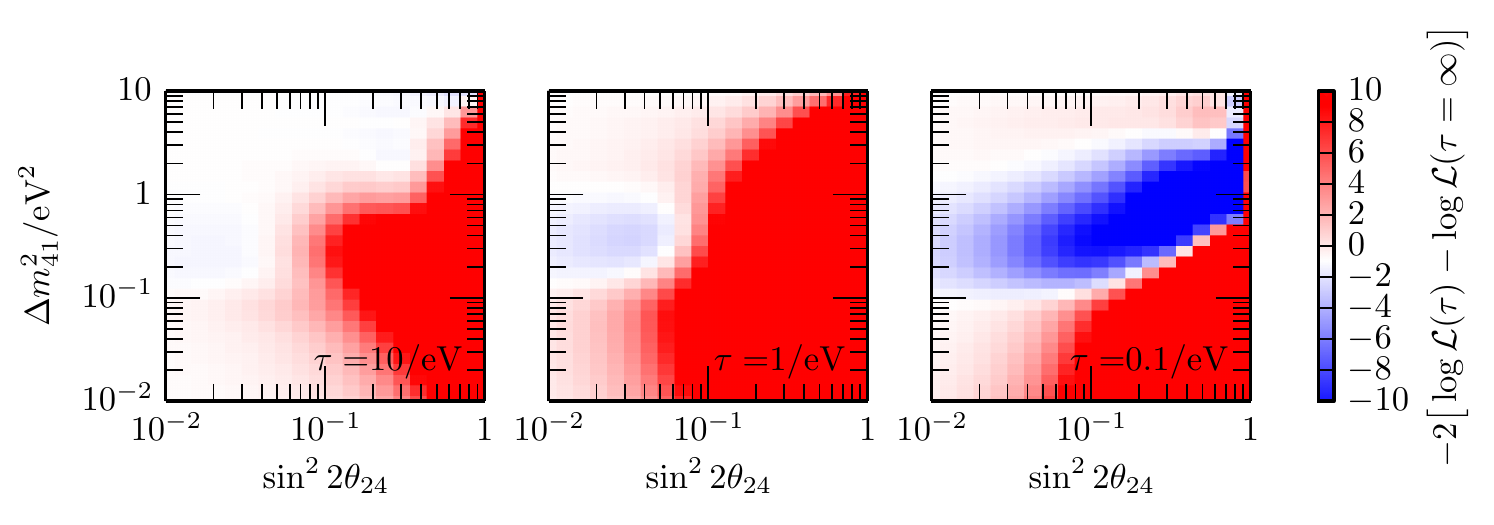}
  \caption{Model comparison between a standard 3+1 model, and a 3+1 model with $\nu_4$ decay, as a function of mixing angle and mass splitting, for three lifetimes. In each $[\Delta m^2_{24}, \sin^2 2 \theta_{24}]$ bin, the likelihood for each model is evaluated using the given sterile mixing angle and mass. Plots from left to right show model comparisons for decay lifetimes $\tau = $ 10, 1, and 0.1 eV$^{-1}$. In all these plots, we assume the decay channel $\nu_4 \rightarrow \nu_3 \phi$. For $\tau$ = 1 eV$^{-1}$, $\hbar c \tau \approx 0.2 \:\mu$m.}
    \label{fig:chi2_diff}
\end{figure*}

The oscillograms shown in Fig.~\ref{fig:oscillograms} indicate that introducing neutrino decay diminishes the strength of the sterile neutrino effect. In order to compare how much the model power changes when we introduce decay, we compare the profile likelihood with and without decay for the scenarios discussed in the previous section. The difference in profile likelihood as a function of sterile neutrino parameters is shown in Fig.~\ref{fig:chi2_diff}. In this figure, red colors indicate that the no-decay scenario is preferred whereas blue colors show preference for the decay solution. As the lifetime decreases, the decay scenario is preferred over the no-decay scenario for $\Delta m^2_{41} \sim 0.1 - 1~{\rm eV^2}$. It is worth noting that the saturated blue color does not necessarily indicate a good fit to data.

\begin{figure*}
  \includegraphics[width=\textwidth]{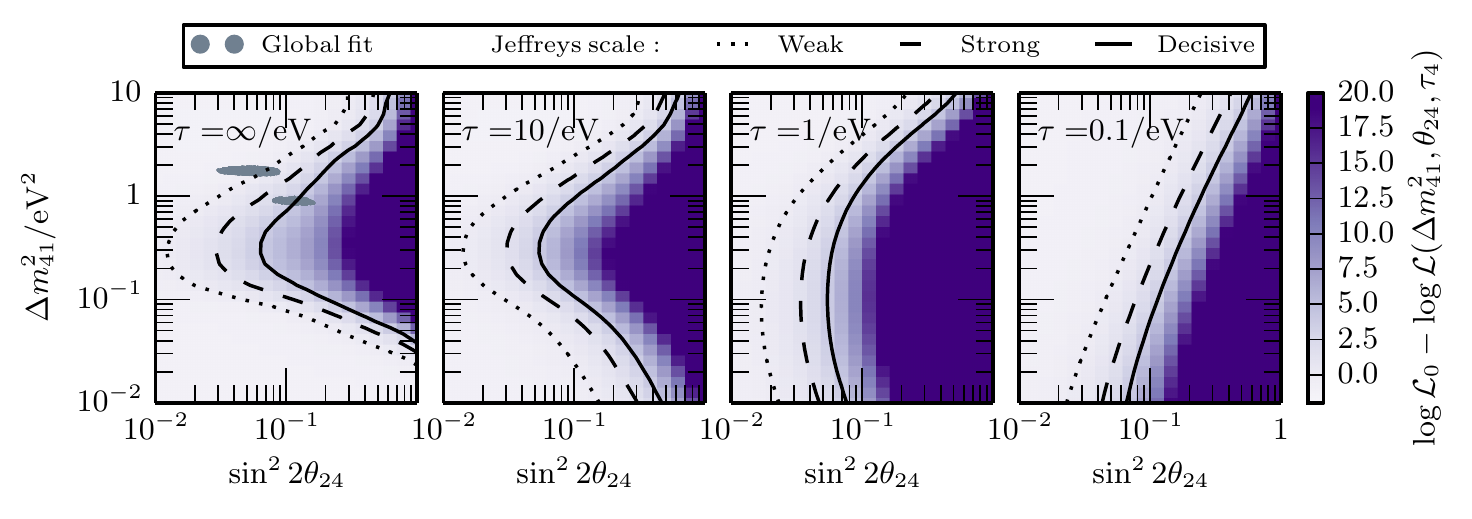}
  \caption{Model comparison between a 3+1 model with decay and the no sterile neutrino scenario as a function of mixing angle and mass, for three lifetimes. Plots from left to right show model comparisons for decay lifetimes $\tau = \infty$ 10, 1, and 0.1 eV$^{-1}$. The dotted, dashed, and solid lines correspond to the Jeffreys scale criterion for weak, strong, and decisive evidence against the alternative hypothesis. In the left-most plot, the grey regions are the 90\% confidence level credible intervals from global fits to short-baseline data \cite{Collin:2016rao}. In all these plots we assume the decay channel $\nu_4 \rightarrow \nu_3 \phi$. For $\tau$ = 1 eV$^{-1}$, $\hbar c \tau \approx 0.2 \:\mu$m.}
    \label{fig:result}
\end{figure*}

Our main result is illustrated in Fig.~\ref{fig:result}, which compares the sterile neutrino hypothesis with decay to the standard three neutrino scenario. In order to quantify the difference between models, we used the approximate Bayes factor, $B_{01}$, as a function of the sterile mixing angle and mass difference for the lifetimes discussed in this paper. The Bayes factor is approximate as we have used a profile instead of a marginal likelihood. In comparing hypotheses, we use the Jeffreys scale, where $ 0 < \log B_{01} <1$, $ 1 < \log B_{01} <2.5 $, and $ 2.5 < \log B_{01} <5 $ correspond to weak, strong, and decisive evidence against a hypothesis, respectively \cite{Trotta:2017wnx}. We observe that for lifetimes on the order of $0.1~{\rm eV}^{-1}$, in the regions of parameter space that correspond to the allowed regions from sterile neutrino global fits to short-baseline data assuming no decay, the sterile neutrino hypothesis is not disfavored. For lifetimes greater than $1~{\rm eV}^{-1}$, the IceCube exclusion is robust under this new physics scenario.

\section{Conclusions\label{sec:conclusions}}

We have reviewed the framework of neutrino decay with oscillations and presented it in a consistent manner using the density matrix formalism. We have further implemented this new physics scenario in the nuSQuIDS software package~\cite{nusquidsdecay}. We have implemented the high-energy IceCube sterile analysis and then introduced the decay of $\nu_4$ as an additional effect. We show that for small values of the lifetime, $\tau \lesssim 0.1~{\rm eV}^{-1}$, the IceCube results interpretation can be significantly changed. Thus, neutrino decay can dramatically alter the landscape of eV-sterile neutrinos and will need to be studied in the context of global fits.

\begin{center}
{ \textbf{Acknowledgments}}
\end{center}

The authors are supported by U.S. National Science Foundation grants 1505858 and 1505855. We would also like to thank Sergio Palomares-Ruiz, Alberto Gago, Christopher Weaver, and Tianlu Yuan for carefully reading this draft and providing us very useful comments. Finally, we would like to thank Jean DeMerit for carefully proofreading our work. 

\pagebreak

\bibliographystyle{apsrev}
\bibliography{nudecay}

\newpage


\appendix
\section{nuSQuIDS decay implementation}\label{App:nuSQuIDSDecay}

{\ttf nuSQuIDS}~\cite{nusquids} is a {\ttf C++} package that calculates the evolution of an ensemble of neutrinos considering oscillations as well as neutrino scattering. Thus it is an easily extendable toolbox for neutrino oscillation experiments and neutrino telescopes. It is written on top of the {\ttf SQuIDS} library~\cite{Delgado:2014kpa}, which implements the density matrix formalism in a numerically optimal way.

In this work we have implemented a {\ttf nuSQuIDS}-derived class that incorporates neutrino decay as discussed in the main text, and can be obtained from~\cite{nusquidsdecay}. We make two additions to the {\ttf nuSQuIDS} virtual functions that govern neutrino noncoherent losses and ensemble interactions: \lstinline[language=c++]{nuSQUIDS::GammaRho} and \lstinline[language=c++]{nuSQUIDS::InteractionsRho}. To the first function we have added a term as given by~\eqref{eq:decay_op} and to the second function we have added $\mathcal{R}$ as given in equations \eqref{eq:r_cpp} and \eqref{eq:r_cvp}.

\subsection{Constructors}

The class specialization is called {\ttf nuSQUIDSDecay}. The main constructor, or ``coupling constructor'' has the following signature
\begin{lstlisting}
nuSQUIDSDecay(marray<double, 1> e_nodes,
              unsigned int numneu,
              NeutrinoType NT,
              bool iinteraction,
              bool decay_regen,
              bool pscalar,
              std::vector<double> m_nu,
              gsl_matrix* couplings),
\end{lstlisting}
where {\ttfamily e\_nodes} is a one dimensional array that gives the energy nodes, {\ttfamily numneu} specifies the number of neutrino states, {\ttfamily iinteraction} toggles neutrino scattering with matter due to DIS interactions, {\ttfamily pscalar} toggles scalar or pseudoscalar couplings (as in the paper, the code assumes either purely scalar or purely pseudoscalar couplings), {\ttfamily decay\_regen} toggles visible and invisible decay inclusion, {\ttfamily m\_nu} is a vector that contains the absolute neutrino masses, and {\ttfamily couplings} is a square real matrix containing $g^{s}_{ij}$ or $g^{p}_{ij}$ as specified by the {\ttfamily pscalar} bool. The code calculates the decay rate matrices according to equations given in Sec.~\ref{sec:decay_rates} from the provided couplings. As a consequence, this constructor has the assumption that the neutrinos are Majorana ``baked-in.'' 

If the user wishes to describe Dirac neutrinos with a different Lagrangian, they can compute their own decay rates and supply them to the following, alternative ``partial rate'' constructor. Its signature is as follows
\begin{lstlisting}
nuSQUIDSDecay(marray<double, 1> e_nodes,
              unsigned int numneu,
              NeutrinoType NT,
              bool iinteraction,
              bool decay_regen,
              bool pscalar,
              bool majorana,
              std::vector<double> m_nu,
              gsl_matrix* rate_matrices[2]).
\end{lstlisting}
The variables with names matching those in the other constructor have the same meanings, but in this version the user can specify the partial widths, $\Gamma_{ij}$ in the rest frame of the decay, instead of the couplings. These should be provided in {\ttfamily rate\_matrices} as an array of two square matrices where the index corresponds to the CPP or CVP process, respectively. The {\ttfamily pscalar} boolean, as before, determines whether these rates are interpreted as scalar or pseudoscalar decay rates. For example, if \lstinline[language=c++]{pscalar=true} then {\ttfamily rate\_matrices[0]}=$\Gamma^{{\rm CPP},p}_{ij}$. The code assumes that the user has calculated these matrices properly. For this reason, if the user is assuming a Majorana neutrino with the Lagrangian given in Eq.~\eqref{eq:int_lagrangian} the use of the coupling constructor is encouraged to minimize the probability of errors.

\subsection{Public members}

The class inherits all public members of the {\ttf nuSQuIDS} class. The most important among them are {\ttfamily EvolveState}, which performs the calculation and {\ttfamily EvalFlavor}, which returns the flavor content. We will not discuss these functions since they are better described in the {\ttf nuSQuIDS} manual. The new public member is
\begin{itemize}
\item \lstinline[language=c++]{Set_DecayRegeneration(bool opt)}: if set to false sets $\mathcal{R} \equiv 0$. This removes the effects of visible decay, reducing the simulation to the invisible decay scenario.
\end{itemize}

\subsection{Provided examples}

Two examples are provided with the code. The first one, {\ttfamily couplings\_example.cpp}, uses the coupling constructor and the second one, {\ttfamily partial\_rate\_example.cpp}, uses the rate constructor. Both the examples calculate the oscillograms corresponding to atmospheric neutrino oscillations. In order to run the examples, we provide the PolyGonato\_QGSJET-II-04 atmospheric neutrino flux calculated with~\cite{mceq}. The {\ttfamily nuSQUIDSDecay} class documentation is included in the code release.

\end{document}